\documentclass[trackchanges,twocolumn]{aastex701}
\DeclareUnicodeCharacter{02BC}{\-}
\usepackage{textcomp, gensymb}
\usepackage{booktabs}
\usepackage{soul}
\usepackage{float}
\usepackage{comment}
\usepackage{array}
\usepackage{makecell}
\usepackage{amsmath}
\usepackage{multirow} 
\usepackage{enumitem}

\defcitealias{Yan_2026}{Y26}

\begin{document}

\title{Distinct Jet Properties in the X-Ray-Obscured State of GRS\,1915+105}

\correspondingauthor{Lang Cui}
\email{cuilang@xao.ac.cn}

\author[0009-0003-6680-1628]{Xi Yan}
\affiliation{State Key Laboratory of Radio Astronomy and Technology, Xinjiang Astronomical Observatory, CAS, 150 Science 1-Street, Urumqi 830011, People's Republic of China}
\email{yanxi@xao.ac.cn}

\author[0000-0003-0721-5509]{Lang Cui}
\affiliation{State Key Laboratory of Radio Astronomy and Technology, Xinjiang Astronomical Observatory, CAS, 150 Science 1-Street, Urumqi 830011, People's Republic of China}
\affiliation{Xinjiang Key Laboratory of Radio Astrophysics, 150 Science 1-Street, Urumqi, Xinjiang 830011, People's Republic of China}
\email[]{cuilang@xao.ac.cn}

\author[0000-0002-7586-5856]{Sergei Trushkin}
\affiliation{Special Astrophysical Observatory of the Russian Academy of Sciences, Nizhny Arkhyz, 369167, Russia}
\email[]{sergei.trushkin@gmail.com}   

\author[0000-0003-2953-6442]{Shuangjing Xu} 
\affiliation{Korea Astronomy and Space Science Institute, 776 Daedeok-daero, Yuseong-gu, Daejeon 34055, Republic of Korea}
\affiliation{Shanghai Astronomical Observatory, Chinese Academy of Sciences, 80 Nandan Road, Shanghai 200030, People's Republic of China}
\email[]{sjxu@kasi.re.kr}   

\author[0000-0001-7369-3539]{Wu Jiang}
\affiliation{Shanghai Astronomical Observatory, Chinese Academy of Sciences, 80 Nandan Road, Shanghai 200030, People's Republic of China}
\email[]{jiangwu@shao.ac.cn}

\author[0000-0002-5385-9586]{Zhen Yan}
\affiliation{Shanghai Astronomical Observatory, Chinese Academy of Sciences, 80 Nandan Road, Shanghai 200030, People's Republic of China}
\email[]{zyan@shao.ac.cn}    

\author[0000-0003-3079-1889]{S\'andor Frey}
\affiliation{Konkoly Observatory, HUN-REN Research Centre for Astronomy and Earth Sciences, Konkoly Thege Mikl\'os \'ut 15-17, 1121 Budapest, Hungary}
\affiliation{CSFK, MTA Centre of Excellence, Konkoly Thege Mikl\'os \'ut 15-17, 1121 Budapest, Hungary}
\affiliation{Department of Astronomy, Institute of Physics and Astronomy, ELTE Eötvös Loránd University, P\'azm\'any P\'eter s\'et\'any 1/A, 1117 Budapest, Hungary}
\email{frey.sandor@csfk.org}

\author[0000-0001-9984-127X]{Timur Mufakharov}
\affiliation{State Key Laboratory of Radio Astronomy and Technology, Xinjiang Astronomical Observatory, CAS, 150 Science 1-Street, Urumqi 830011, People's Republic of China}
\affiliation{Special Astrophysical Observatory of the Russian Academy of Sciences, Nizhny Arkhyz, 369167, Russia}
\email[]{timur.mufakharov@gmail.com}   

\author[0000-0002-1404-8924]{Ruchika Dhaka}
%\affiliation{Department of Physics, IIT Kanpur, Kanpur, Uttar Pradesh 208016, India}
\affiliation{State Key Laboratory of Radio Astronomy and Technology, Xinjiang Astronomical Observatory, CAS, 150 Science 1-Street, Urumqi 830011, People's Republic of China}
\email[]{ruchikadhaka1997@gmail.com}

\begin{abstract}

GRS\,1915+105 has remained in an X-ray-obscured state since its transition from a long-lasting unobscured state in 2019. We report on 6.7-GHz East Asia VLBI Network observations of GRS\,1915+105 obtained during strong radio flares detected at 2.3--11.2\,GHz with the RATAN-600 radio telescope in 2025. Our images reveal two contrasting jet morphologies. The first epoch, associated with a flare evolving from an optically thick to an optically thin spectrum, shows a bright radio core accompanied by an extended jet structure. By contrast, the second epoch, observed near the peak of another flare displaying optically thin emission at lower frequencies, is dominated by two bright, symmetric, well-separated jet blobs and shows no detectable radio core. If these jets exhibited the apparent superluminal motions commonly observed prior to 2019, measurable angular shifts would be expected over the five-hour observations. However, no significant jet motion is detected. Combined with our derived jet speed of $\beta\Gamma \lesssim 0.40$, these results suggest that the jets launched during the current obscured state are slower than the relativistic jets ($\beta\Gamma \gtrsim 1$) observed earlier during the unobscured state. Together with the recently reported large variations in jet orientation, our findings in GRS\,1915+105 provide robust support for the emerging paradigm that X-ray binary jets launched in obscured and unobscured states likely exhibit distinct propagation properties. 

\end{abstract}


\section{Introduction} \label{sec:Introduction}
GRS\,1915+105 was the first Galactic X-ray binary (XRB) in which apparent superluminal jets were observed \citep{Mirabel_1994Natur.371...46M,Mirabel_1999ARA&A..37..409M}. It has since become an archetypal black hole X-ray binary for studying the connection between accretion states and the formation of relativistic jets through coordinated X-ray, infrared, and radio observations \citep[e.g.,][]{Fender_2004ARA&A..42..317F}. However, progress in understanding the accretion--jet coupling in this source has been severely limited since 2019, when GRS\,1915+105 transitioned from a long-lasting unobscured state into an X-ray-obscured state \citep[e.g.,][]{Miller_2020ApJ...904...30M,Miller_2025ApJ...995L..14M,Motta_2021MNRAS.503..152M,Gandhi_2025MNRAS.537.1385G}. This obscured state is characterized by strong local absorption, with hydrogen column densities of $N_{\rm H}\sim10^{22}$--$10^{24}\,\rm cm^{-2}$ \citep[e.g.,][]{Balakrishnan_2021ApJ...909...41B,Athulya_2023MNRAS.525..489A}. Despite the lack of direct access to the central accretion flow, the source continues to exhibit episodic strong radio flares \citep[e.g.,][]{Motta_2021MNRAS.503..152M,Trushkin_2023ATel16168....1T,Trushkin_2023ATel15964....1T,Trushkin_2023ATel15974....1T,Trushkin_2025ATel16976....1T}, indicating the jet activities. Notably, the latest MeerKAT monitoring in June 2026 suggests that GRS\,1915+105 has reached its lowest radio luminosity since its discovery \citep{Motta_2026ATel17865....1M}.

Recently, \citet{Fender_2025NatAs...9.1854F,fender2026linkobscuredaccretionmildly} established a new paradigm for understanding jet formation in XRBs based on a statistically significant sample. Their results suggest that fast relativistic jets (with Lorentz factor $\Gamma >2$) are preferentially produced in unobscured environments and remain aligned with a stable axis, likely associated with the black hole spin. In contrast, jets launched in highly obscured systems tend to be slower, with typical speeds of $\sim 0.3\,c$ (here $c$ denotes the speed of light), and are likely to undergo precession. 

The observed differences in the jet properties of GRS\,1915+105 are broadly consistent with this emerging picture. Prior to 2019, when the source was in the unobscured state, the jets were consistently observed along the southeast--northwest direction with a mean position angle (PA) of $147^{\circ}\pm8^{\circ}$ \citep{Rodr_2025ApJ...986..108R}. Kinematic studies also suggest that the jets exhibited relativistic bulk motions with $\beta\Gamma \gtrsim 1$, where $\beta$ is the intrinsic jet speed in units of $c$ \citep[e.g.,][]{Mirabel_1994Natur.371...46M,Rodr_1999ApJ...511..398R,Fender_1999MNRAS.304..865F,Miller-Jones_2005MNRAS.363..867M,Miller-Jones_2007MNRAS.375.1087M}. 

However, observations obtained during the obscured state reveal markedly different jet properties. \citet{Rodr_2025ApJ...986..108R} were the first to report significant changes in both the jet PA and viewing angle relative to their historical values, by $24^\circ$ and $17^\circ$, respectively. \citet{Jiang_2026ApJ..1000L..45J} also observed an unusual north--south jet orientation ($\mathrm{PA}=188^\circ\pm3^\circ$) and measured a much lower jet speed of $\beta\Gamma = 0.37$. In \citet{Yan_2026} (hereafter \citetalias{Yan_2026}), we found that the jet orientation has undergone substantial variations since 2023 ($\mathrm{PA}=118^\circ$--$188^\circ$), which may be indicative of jet precession. These results collectively suggest that the kinematic properties and orientation of the jets of GRS\,1915+105 in the obscured state likely differ from those of the pre-2019 jets observed during the unobscured state. Despite these indications, the post-2019 jets remain poorly understood compared to the well-studied relativistic jets launched earlier.

Building on the very long baseline interferometry (VLBI) imaging results presented in \citetalias{Yan_2026}, we perform a detailed analysis of the physical properties of GRS\,1915+105 jets observed in 2025 (during the obscured state) and place them in the context of the well-studied pre-2019 relativistic jets.
In Section~\ref{sec:Observations}, we describe the data analysis. The results are presented in Section~\ref{sec:Results}, followed by a discussion in Section~\ref{sec:Discussion}. Finally, our main conclusions are summarized in Section~\ref{sec:Summary}. Throughout this paper, we adopt a distance of $9.4\pm1.0$\,kpc for GRS\,1915+105 \citep{Reid_2023ApJ...959...85R}.

\begin{figure*}[t!]
\centering
\includegraphics[width=0.46\linewidth]{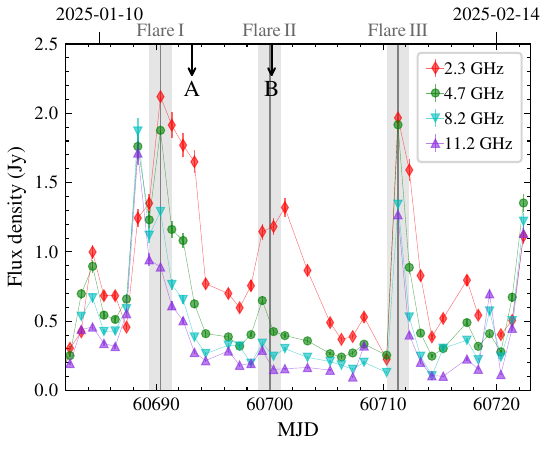}
\hspace{5cm}
\includegraphics[width=0.46\linewidth]{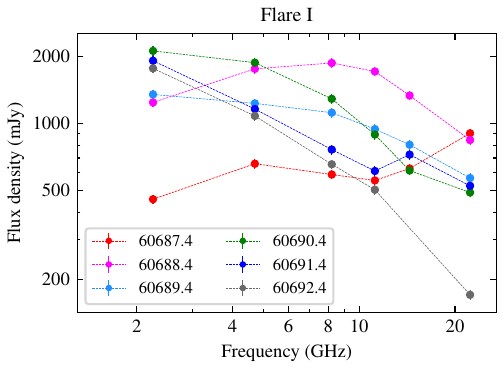}
%\hspace{0.5cm}
\includegraphics[width=0.46\linewidth]{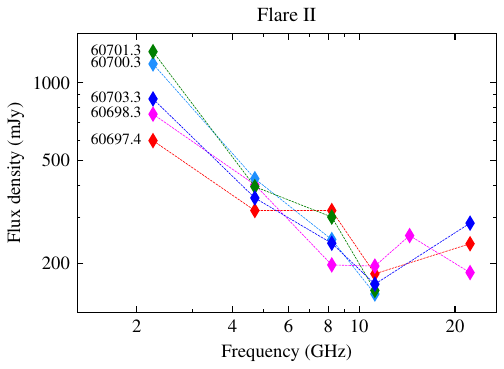}
\caption{
Top panel: radio light curves of GRS\,1915+105 from January to February 2025, observed with RATAN at 2.3, 4.7, 8.2, and 11.2\,GHz. The gray shaded regions indicate three major radio flares, occurring at MJD $60690.4 \pm 1$ (Flare~I), $60700.0 \pm 1$ (Flare~II), and $60711.3 \pm 1$ (Flare~III). The EAVN observing epochs (epoch~A at MJD~60693.1 and epoch~B at MJD~60700.2) are marked on the top axis. Bottom panels: radio spectral evolution of GRS\,1915+105 during Flare~I and Flare~II, based on RATAN observations at 2.3, 4.7, 8.2, 11.2, 14.4, and 22.3\,GHz, with all frequencies measured within 1\,min.
}
\label{fig:GRS1915_RATAN_LC} 
\end{figure*}

\begin{figure*}[htbp!]
\begin{center}
    \centering
    \hspace{-0.5cm}
    \includegraphics[width=0.34\linewidth]{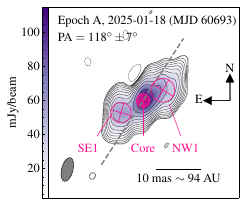}
    \hspace{1.05cm}
    \includegraphics[width=0.33\linewidth]{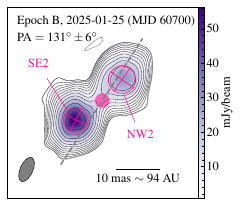}
    \includegraphics[width=0.35\linewidth]{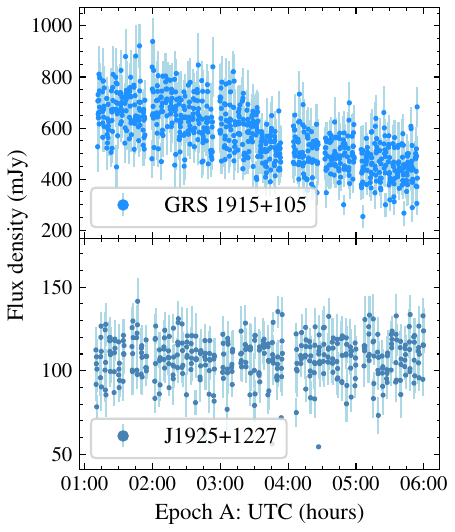}
    \hspace{1cm}
    \includegraphics[width=0.35\linewidth]{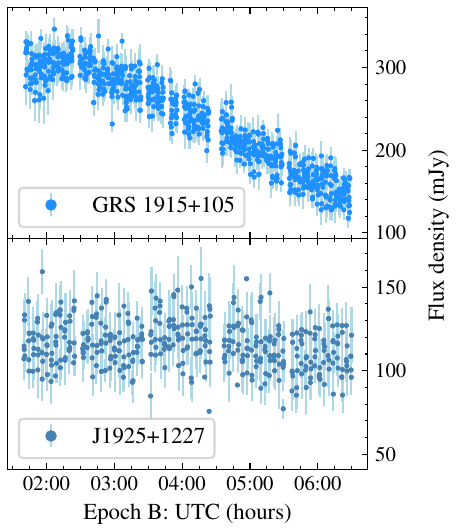}
    \caption{
    \textit{Top:} EAVN 6.7-GHz images of GRS\,1915+105 obtained in 2025. Contours start at 2.7\,mJy\,beam$^{-1}$ (epoch~A) and 0.9\,mJy\,beam$^{-1}$ (epoch~B), and increase by successive factors of 2. Negative contours are shown with dashed lines. 
    The elliptical Gaussian synthesized beam sizes (half-power widths) are $5.50\,\mathrm{mas} \times 2.60\,\mathrm{mas}$ at major axis position angle of PA $= -14\fdg3$ (epoch~A) and $5.87\,\mathrm{mas} \times 2.87\,\mathrm{mas}$ at PA $= -24\fdg3$ (epoch~B), as indicated in the lower-left corner of each panel. 
    Open magenta circles mark the positions and sizes of the model-fitted Gaussian components. Filled magenta circles indicate the core position derived from epoch~A \citepalias[see][]{Yan_2026}. The jet PA, measured by \citetalias{Yan_2026}, is also indicated. The gray dashed lines represent the typical southeast--northwest jet orientation with a mean PA of $147^{\circ}\pm8^{\circ}$ \citep{Rodr_2025ApJ...986..108R}. 
    \textit{Bottom panels:} VLBI flux density as a function of time for the target GRS\,1915+105 (upper panel) and the calibrator J1925+1227 (lower panel). The time resolution is 10\,s.
    }
    \label{fig:GRS1915_images_and_VLBI_LC} 
\end{center}
\end{figure*}

\section{Observations and Data Analysis} \label{sec:Observations}
During the giant radio flares of GRS\,1915+105 in early 2025 \citep{Trushkin_2025ATel16976....1T}, we observed the source with East Asia VLBI Network (EAVN)\footnote{EAVN: \url{https://radio.kasi.re.kr/eavn/main.php}} at 6.7\,GHz. The observations were conducted over six epochs, each with a duration of about 5 hours. However, reliable fringes were not detected in three epochs, while another epoch has already been discussed in detail by \citetalias{Yan_2026}. In this work, we focus on the remaining two epochs (A and B), which provide the high-quality data for probing the physical properties of the jets. For details of the observations and data reduction, we refer to \citetalias{Yan_2026}.

To search for rapid variability on hour timescales, we extracted the VLBI flux density directly from the calibrated visibility data in the Astronomical Image Processing System \citep[{\tt AIPS};][]{Greisen2003}. This was performed using the {\tt DFTPL} task with a time bin of 10\,s. As to be shown in Section~\ref{Results:jet_morphology_and_LC}, GRS\,1915+105 is well-resolved by our observations, resulting in significantly different flux density variations on different baselines. We note that the shortest baselines in epochs~A and B have lengths of about 400 and 300\,km, corresponding to angular resolutions of approximately 23 and 31\,mas, respectively. At these resolutions, the source is expected to remain largely unresolved. We therefore extracted the flux density from the shortest baseline in each epoch, which could reflect the intrinsic source variability. For comparison, we also extracted light curves of the calibrator ICRF J192540.8+122738 (hereafter J1925+1227) from the short baselines.

\begin{figure*}[htbp!]
\centering
\begin{minipage}[c]{0.26\textwidth}
    \centering
    \includegraphics[width=\linewidth]{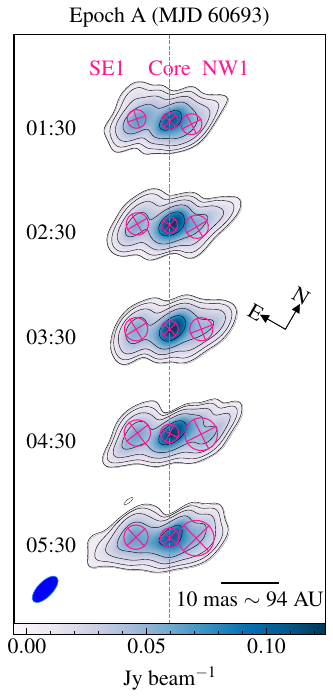}
\end{minipage}
\hspace{1cm}
\begin{minipage}[c]{0.36\textwidth}
    \centering
    \includegraphics[width=1\linewidth]{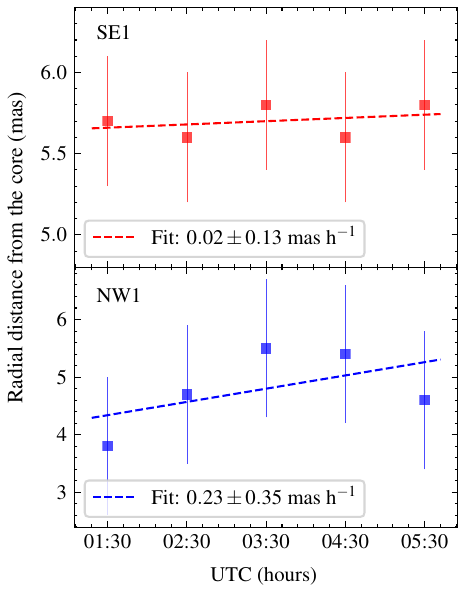}
\end{minipage}
\caption{
\textit{Left:} time-binned EAVN 6.7-GHz images of GRS\,1915+105 from epoch~A, observed on 2025 January 18 (MJD~60693). All images are restored with a common synthesized beam of $5.82\,\mathrm{mas} \times 2.64\,\mathrm{mas}$ at $\mathrm{PA}=-14^\circ$ and rotated clockwise by $30^\circ$ to facilitate comparison. Contours start at 0.005\,Jy\,beam$^{-1}$ and increase by successive factors of 2. The magenta circles indicate the positions and sizes of the fitted Gaussian components. The images are aligned on the identified core position.
\textit{Right:} radial distance from the core as a function of time for the SE1 and NW1 components. The dashed lines show the linear fits used to derive the apparent proper motions.
}
\label{fig:GRS1915_epoch_A}
\end{figure*}

\begin{figure*}[htbp!]
\centering
\begin{minipage}[c]{0.26\textwidth}
    \centering
    \includegraphics[width=\linewidth]{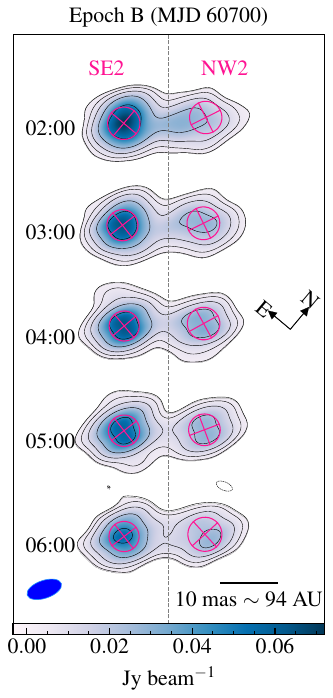}
\end{minipage}
\hspace{1cm}
\begin{minipage}[c]{0.36\textwidth}
    \centering
    \includegraphics[width=1\linewidth]{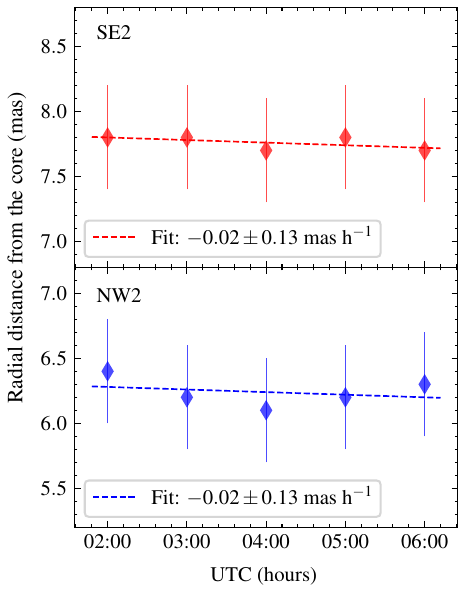}
\end{minipage}
\caption{
\textit{Left:} time-binned EAVN 6.7-GHz images of GRS\,1915+105 from epoch~B, observed on 2025 January 25 (MJD~60700). All images are restored with a common synthesized beam of $6.24\,\mathrm{mas} \times 3.07\,\mathrm{mas}$ at $\mathrm{PA}=-31^\circ$ and rotated clockwise by $40^\circ$ to facilitate comparison. Contours start at 0.003\,Jy\,beam$^{-1}$ and increase by successive factors of 2. The magenta circles indicate the positions and sizes of the fitted Gaussian components. The images are aligned on the core position inferred from epoch~A.
\textit{Right:} radial distance from the core as a function of time for the SE2 and NW2 components. The dashed lines show the linear fits used to derive the apparent proper motions.
}
\label{fig:GRS1915_epoch_B}
\end{figure*}

\section{Results} \label{sec:Results}
\subsection{Jet Morphologies and VLBI Flux Density Variability} 
\label{Results:jet_morphology_and_LC}
In the top panel of \autoref{fig:GRS1915_RATAN_LC}, we present the radio light curves of GRS\,1915+105 obtained with the RATAN-600 radio telescope \citep[hereafter RATAN;][]{Trushkin_2000A&AT...19..525T,Trushkin_2008mqw..confE..32T} at 2.3--11.2\,GHz between January and February 2025. Three major flares (see details in \citetalias{Yan_2026}) and the EAVN epochs are also indicated in the figure. As shown, the epoch~A was conducted about three days after Flare~I, while epoch~B almost coincided with the peak of Flare~II. 

The bottom panels show the radio spectral evolution of GRS\,1915+105 during Flare~I and Flare~II, based on RATAN observations over the frequency range of 2.3--22.3\,GHz. The flux densities at all frequencies were measured almost simultaneously within 1\,min. On MJD~60687, we measured a spectral index ($\alpha = \log(S_{2}/S_{1}) / \log(\nu_{2}/\nu_{1})$) of $\alpha = 0.30 \pm 0.16$. On MJD~60688, the spectrum exhibits optically thick emission ($\alpha = 0.32 \pm 0.09$) at $\nu \leq 8.2$\,GHz and optically thin emission ($\alpha = -0.80 \pm 0.07$) at $\nu \geq 8.2$\,GHz. From MJD~60689 to 60692, the spectra remain optically thin over the entire 2.3--22.3\,GHz frequency range, with $\alpha$ varying from $-0.38 \pm 0.16$ to $-1.02 \pm 0.16$. These results indicate that Flare~I evolved from an optically thick spectrum during the rise phase to an optically thin spectrum during the decay phase. During Flare~II (MJDs~60697--60703), we observed steep optically thin spectra with $\alpha \approx -1$ at $\nu \leq 11.2$\,GHz. Interestingly, the spectra become nearly flat or inverted at $\nu \geq 11.2$\,GHz.

The top panels of \autoref{fig:GRS1915_images_and_VLBI_LC} show the EAVN 6.7-GHz images of GRS\,1915+105. As reported in \citetalias{Yan_2026}, the radio core is located at the map peak in epoch~A and within the emission gap between the two jet components in epoch~B. The source structures in epochs~A and B were modeled using several circular Gaussian components (see Table~2 of \citetalias{Yan_2026}), which are overlaid on the images.

Epoch~A, associated with Flare~I, reveals a bright radio core accompanied by extended jet emission on both sides. The flux densities of the core, SE1, and NW1 components are $186\pm37$, $153\pm31$, and $207\pm41$\,mJy, respectively \citepalias{Yan_2026}. Notably, the receding-side component NW1 appears brighter than the approaching-side component SE1. This is likely caused by blending between NW1 and the core emission, as can be seen in the image. The jet orientation was suggested to be $\mathrm{PA}=118^{\circ}\pm7^{\circ}$ \citepalias{Yan_2026}. Epoch~B, associated with Flare~II, exhibits a markedly different morphology. No core component is robustly detected. Instead, the source structure is dominated by two symmetric, well-separated jet blobs, SE2 and NW2, with flux densities of $163\pm33$ and $104\pm21$\,mJy, respectively \citepalias{Yan_2026}. The jet PA is measured to be $131^{\circ}\pm6^{\circ}$ \citepalias{Yan_2026}.

The bottom panels of \autoref{fig:GRS1915_images_and_VLBI_LC} show the VLBI flux density as a function of time for GRS\,1915+105 and the calibrator J1925+1227. In both epochs, GRS\,1915+105 shows an overall decline in flux density, but the decline is more pronounced in epoch~B than in epoch~A. As a comparison, the light curves of J1925+1227 show no significant variability, with mean flux densities of $107 \pm 11$\,mJy and $116 \pm 13$\,mJy in epochs~A and B, respectively. This confirms that the observed hour-scale variability of GRS\,1915+105 is intrinsic to the source rather than resulting from calibration uncertainties.

In summary, epoch~A was observed during the flux-density decay phase of Flare~I, which evolved from an optically thick to an optically thin spectrum. At this stage, we detected a bright radio core accompanied by an extended two-sided jet structure. Epoch~B was obtained near the peak of Flare~II, whose spectra remained optically thin at $\nu \leq 11.2\,\mathrm{GHz}$ throughout both its rise and decay phases. Unlike epoch~A, the core emission appears to be strongly suppressed; instead, the radio radiation was dominated by two bright, symmetric, well-separated jet blobs. The VLBI light curves in both epochs show a gradual decline in flux density with time. These findings point to a possible connection between the radio spectral evolution and the mas-scale jet morphology.

\subsection{Apparent Jet Motions} \label{Results:jet_motions} 
To measure jet motions, we divided each 5-hour observation in epochs~A and B into five time bins. First, the {\tt AIPS}--calibrated visibility data were split into five segments, which were independently self-calibrated and imaged using the {\tt DIFMAP} package \citep{Shepherd_Difmap_1997ASPC..125...77S}. The source structure in each time bin was then modeled with several circular Gaussian functions to determine the positions of the jet components relative to the core. For epoch~B, where the core is not directly detected, we adopted the core position inferred from epoch~A (see \autoref{fig:GRS1915_images_and_VLBI_LC} and more details in \citetalias{Yan_2026}). Finally, apparent proper motions were derived by fitting a linear function to the radial separation of each component from the core as a function of time.

As shown in \autoref{fig:GRS1915_epoch_A}, the proper motions of the SE1 and NW1 components in epoch~A are $0.02 \pm 0.13$\,mas\,h$^{-1}$ and $0.23 \pm 0.35$\,mas\,h$^{-1}$, respectively, indicating no significant detection of jet motions. The SE1 component remains approximately stationary in radial distance, whereas NW1 shows larger positional variations. We note that NW1 is less well constrained due to partial blending with the core emission, which may introduce additional uncertainties in its measured position and size. \autoref{fig:GRS1915_epoch_B} shows the results for epoch~B. No jet motion is detected in either the SE2 or NW2 component, with apparent proper motions of $-0.02 \pm 0.13$\,mas\,h$^{-1}$. 

Based on the angular resolution of the array, we can place an upper limit on the apparent jet motion (see also discussion in Section~\ref{Discussion:slower_jets}). The synthesized beam has a major and minor axis of approximately 6 and 3\,mas, respectively (see \autoref{fig:GRS1915_images_and_VLBI_LC}). As a conservative estimate, we assume that only positional shifts exceeding one quarter of the major-axis size can be observed, corresponding to a displacement of $\sim$1.5\,mas. The absence of any significant jet motion in either epoch~A or epoch~B suggests that the radial displacement of the jet may be smaller than $\sim$1.5\,mas over the 5-hour observing interval. This corresponds to an upper limit on the proper motion of $\lesssim 0.3$\,mas\,h$^{-1}$. At a distance of 9.4\,kpc, this translates into an apparent jet speed of $\beta_{\rm app} \lesssim 0.39$ (see Eq.~\ref{eq:beta_app} in Section~\ref{Discussion:slower_jets}).

\subsection{Intrinsic Jet Speed} \label{Results:intrinsic_speed}
The absence of detectable jet motions prevents the direct measurement of both the apparent and intrinsic speeds. We therefore adopt an alternative approach based on the assumption that the two-sided jets in GRS\,1915+105 are ejected simultaneously from the core and propagate ballistically. We note that both the recent VLBI and Karl G. Jansky Very Large Array observations obtained during two radio flares in 2023 suggest a jet inclination close to $90^\circ$, for which relativistic boosting and de-boosting effects are expected to be insignificant. Interestingly, the approaching and receding jet blobs exhibit almost the same separations from the core, intrinsic speeds, and flux densities \citep{Rodr_2025ApJ...986..108R,Jiang_2026ApJ..1000L..45J}. Under the assumption of intrinsically symmetric two-sided jets in GRS\,1915+105, the quantity $\beta \cos\theta$ (where $\theta$ is the jet viewing angle) can be constrained following \citet{Rodr_2025ApJ...986..108R}:
\begin{equation}
\beta \cos\theta = \frac{\Delta r_\mathrm{app} - \Delta r_\mathrm{rec}}{\Delta r_\mathrm{app} + \Delta r_\mathrm{rec}},
\end{equation}
where $\Delta r_\mathrm{app}$ and $\Delta r_\mathrm{rec}$ are the angular separations of the approaching and receding components relative to the core, respectively. Therefore, we can estimate the intrinsic jet speed $\beta$ once $\Delta r_\mathrm{app}$, $\Delta r_\mathrm{rec}$, and the jet viewing angle $\theta$ are known.

This method is particularly suitable for epoch~B, where the approaching (SE2) and receding (NW2) blobs are well separated and appear almost symmetric. As shown in \autoref{fig:GRS1915_epoch_B}, we measure $\Delta r_\mathrm{app} = 7.8 \pm 0.4$\,mas and $\Delta r_\mathrm{rec} = 6.2 \pm 0.4$\,mas, yielding $\beta \cos\theta = 0.11 \pm 0.04$.

On the other hand, $\beta\cos{\theta}$ can also be derived from the jet-to-counterjet brightness ratio (or flux-density ratio) and the spectral index, according to
\begin{equation}
\beta\cos{\theta} = \left(\frac{R_{\rm B}^{1/(k-\alpha)}-1}{R_{\rm B}^{1/(k-\alpha)}+1}\right),
\end{equation}
where, theoretically, $k=2$ for continuous jets and $k=3$ for discrete components. This method also assumes that the twin jets are intrinsically symmetric.

\begin{figure}[t!]
\centering
\includegraphics[width=1\linewidth]{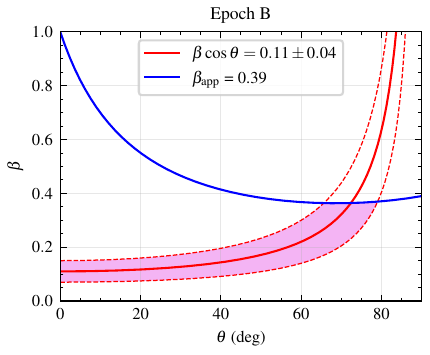}
\caption{
The intrinsic jet speed ($\beta$) as a function of viewing angle ($\theta$). The red curve corresponds to the constraint $\beta\cos\theta = 0.11 \pm 0.04$, with the dashed curves indicating the $1\sigma$ uncertainty. The blue curve shows the upper limit on the intrinsic jet speed derived from the constraint $\beta_{\rm app} \lesssim 0.39$. The allowed parameter space is bounded by these two constraints and is represented by the shaded region.
}
\label{fig:GRS1915_jet_beta_theta} 
\end{figure}

For epoch~B, we measure $\alpha = -0.98 \pm 0.13$ using the RATAN flux densities of $425 \pm 21$ and $247 \pm 12$\,mJy at 4.7 and 8.2\,GHz on MJD~60700, respectively (see \autoref{fig:GRS1915_RATAN_LC}). The flux-density ratio is calculated to be $R_{\rm B} = 1.6 \pm 0.4$ from the flux densities of the SE2 and NW2 components. As for $k$, previous studies have suggested values in the range of 1.3--2.5 \citep{Mirabel_1994Natur.371...46M,Fender_1999MNRAS.304..865F,Miller-Jones_2005MNRAS.363..867M}. Taken together, these measurements yield a $\beta\cos{\theta}$ range from $0.07 \pm 0.04$ to $0.10\pm0.05$. Notably, the two independent methods give consistent constraints on $\beta\cos{\theta}$ for epoch~B.

Using the derived value of $\beta\cos\theta = 0.11 \pm 0.04$ together with the upper limit on the apparent jet speed, $\beta_{\rm app} \lesssim 0.39$ (see Section~\ref{Results:jet_motions}), we can further constrain the intrinsic jet speed for epoch~B. As shown in \autoref{fig:GRS1915_jet_beta_theta}, the measured $\beta\cos\theta$ defines a relation between the intrinsic speed and viewing angle of the jet. Independently, the constraint $\beta_{\rm app} \lesssim 0.39$ provides an upper limit on $\beta$ as a function of $\theta$ (see Eq.~\ref{eq:beta} in Section~\ref{Discussion:slower_jets}). The intersection region of these two constraints yields a viewing angle of $\theta \lesssim 78^\circ$ and an intrinsic jet speed of $\beta \lesssim 0.37$, corresponding to $\beta\Gamma \lesssim 0.40$.

For epoch~A, the jet components are less well separated and NW1 exhibits positional fluctuations, making the estimation of intrinsic jet speed more uncertain. Nevertheless, we obtain a tentative estimate of $\beta \cos\theta = 0.09 \pm 0.12$ by adopting the average separations of $\Delta r_\mathrm{app} = 5.7 \pm 0.4$\,mas (SE1) and $\Delta r_\mathrm{rec} = 4.8 \pm 1.1$\,mas (NW1). Similarly, we also obtain an upper limit of $\beta \lesssim 0.37$ (or $\beta\Gamma \lesssim 0.40$); however, we note that the estimate of $\beta\cos\theta$ for epoch~A is subject to the large uncertainties.

\begin{figure*}[t!]
\centering
\includegraphics[width=0.52\linewidth]{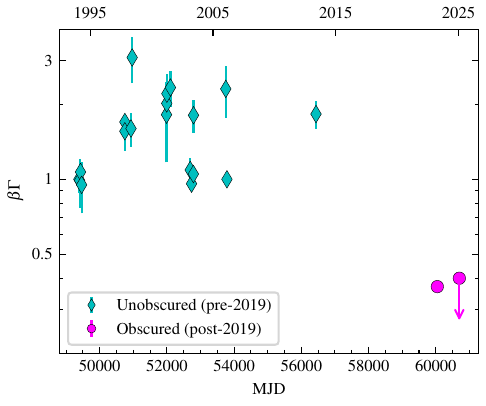}
\caption{Jet speed of GRS\,1915+105 as a function of time. All values are calculated based on a source distance of $9.4 \pm 1.0$\,kpc (see Section~\ref{Discussion:slower_jets}). The magenta points represent measurements from \citet{Jiang_2026ApJ..1000L..45J} and this work (see \autoref{table:comparison}).
}
\label{fig:GRS1915_jet_speed} 
\end{figure*}

\section{Discussion} \label{sec:Discussion}
\subsection{Evidence for Slower Jets in the Obscured State} \label{Discussion:slower_jets}
In this section, we compare the properties of the 2025 jets in the X-ray-obscured state with those of the well-studied pre-2019 relativistic jets in the unobscured state.

The absence of detectable jet motions (see Section~\ref{Results:jet_motions}) suggests that the 2025 jets do not exhibit the large apparent superluminal motions commonly observed for the pre-2019 jets. On VLBI scales, \citet{Dhawan_2000ApJ...543..373D} measured an apparent speed of $0.92 \pm 0.08$\,mas\,h$^{-1}$ (corresponding to $\sim22.1$\,mas\,d$^{-1}$) for the approaching jet of GRS\,1915+105 during 1997--1998. In addition, \citet{Reid_2014} observed a radial offset of $\sim 4.9$\,mas for the approaching jet over a 5-hour interval in 2013, corresponding to a proper motion of $23.6 \pm 0.5$\,mas\,d$^{-1}$. If the approaching jets observed in 2025 moved at a similar apparent speed, a projected displacement of $\sim 5$\,mas would be expected over our 5-hour observing interval. Such a displacement is comparable to the synthesized beam major axis ($5.50$--$5.87$\,mas; see \autoref{fig:GRS1915_images_and_VLBI_LC}) and should therefore be clearly detectable. However, no significant motion is detected. By contrast, \citet{Jiang_2026ApJ..1000L..45J} reported an apparent speed of $6.50 \pm 0.18$\,mas\,d$^{-1}$ for jets produced during the obscured state in 2023. Over a 5-hour interval, this would correspond to a projected shift of only $\sim1.3$\,mas, less than one quarter of the synthesized beam major axis and therefore unlikely to be reliably resolved in our observations.

On the other hand, we note that previous studies reported $0.32 \leq \beta\cos\theta \leq 0.46$ for the pre-2019 jets \citep{Mirabel_1994Natur.371...46M,Fender_1999MNRAS.304..865F,Miller-Jones_2007MNRAS.375.1087M}. However, in this work, we obtain a smaller value of $\beta\cos\theta \approx 0.11$ for epoch B and derive a jet speed of $\beta\Gamma \lesssim 0.40$ (see Section~\ref{Results:intrinsic_speed}). This value is consistent with the recent measurement of $\beta\Gamma = 0.37$ for the symmetric ejecta observed in 2023, with an unusual orientation at $\mathrm{PA}=188^\circ\pm3^\circ$ \citep{Jiang_2026ApJ..1000L..45J}.

To investigate the long-term evolution of the jet speed in GRS\,1915+105, we derived historical values of $\beta$ for the pre-2019 jets using literature measurements of the apparent proper motions of the approaching and receding jets ($\mu_{\rm app}$ and $\mu_{\rm rec}$), compiled in Table~3 of \citetalias{Yan_2026}. When both $\mu_{\rm app}$ and $\mu_{\rm rec}$ are available, the jet viewing angle can be determined based on a distance of $d = 9.4 \pm 1.0$\,kpc for GRS\,1915+105 (see \citetalias{Yan_2026}), allowing the intrinsic jet speed to be calculated using \citep{Mirabel_1994Natur.371...46M}:
\begin{equation}
\beta = \frac{1}{\cos\theta }\frac{\mu_{\rm app} - \mu_{\rm rec}}{\mu_{\rm app} + \mu_{\rm rec}} .
\end{equation}

On the other hand, for observations in which only $\mu_{\rm app}$ is available, we derived the intrinsic jet speed using:
\begin{equation} \label{eq:beta_app}
\beta_{\rm app} = \frac{1}{173} \frac{\mu_{\rm app}}{\rm mas\,d^{-1}} \frac{d}{\rm kpc},
\end{equation}
and
\begin{equation} \label{eq:beta}
\beta = \frac{\beta_{\rm app}}{\sin\theta + \beta_{\rm app}\cos\theta}.
\end{equation}
For these calculations, we adopted a weighted mean viewing angle of $\theta = 64^\circ \pm 4^\circ$ for the pre-2019 jets reported by \citet{Reid_2023ApJ...959...85R}.

\begin{deluxetable*}{cccccc}[htbp!]
\renewcommand\arraystretch{1.1}
\tabletypesize{\footnotesize}
\tablecaption{Comparison of Jet Properties in the Obscured and Unobscured States of GRS\,1915+105
\label{table:comparison}}
\tablehead{
\colhead{} &
\multicolumn{2}{c}{January 2025} &
\colhead{Pre-2019}\\
\cmidrule(lr){2-3}
&
\colhead{Epoch A} &
\colhead{Epoch B} &
\colhead{Typical values}
}
\startdata
Associated flare & \makecell{Three days after Flare~I \\ (optically thick $\rightarrow$ thin)} & \makecell{Near the peak of Flare~II \\ (optically thin at $\nu \leq 11.2$\,GHz)} & ... \\
Hour-scale variability & Overall decline in flux density & Overall decline in flux density & ... \\
Jet morphology & \makecell{Bright core with an extended two-sided jet} & \makecell{Two symmetric jet blobs \\ with no detectable radio core} & ... \\
\hline
Obscured state? & Yes & Yes & No \\
Jet orientation & $118^\circ \pm 7^\circ$ & $131^\circ \pm 6^\circ$ & \makecell{$130^\circ$--$160^\circ$, \\ $147^\circ \pm 8^\circ$ (mean)} \\
$\beta\cos\theta$ & $0.09 \pm 0.12$ & $0.11 \pm 0.04$ & $0.32$ -- $0.46$ \\
$\beta\Gamma$ & $\lesssim 0.40$ & $\lesssim 0.40$ & $\gtrsim 1$\\
\enddata
\end{deluxetable*}

\autoref{fig:GRS1915_jet_speed} shows the jet speed of GRS\,1915+105 as a function of time. While the pre-2019 jets observed during the unobscured state are consistently relativistic ($\beta\Gamma \sim 1$--3), the post-2019 jets launched during the obscured state are characterized by smaller values ($\beta\Gamma \lesssim 0.40$). 
Interestingly, \citetalias{Yan_2026} also suggested that, unlike the pre-2019 jets that were broadly aligned at relatively stable axis, the post-2019 jets exhibit large variations in orientation, with a PA range of $118^{\circ}$--$188^{\circ}$.

In summary, our results suggest that jets launched during the obscured state of GRS\,1915+105 differ from the relativistic outflows observed during the unobscured state, exhibiting lower speeds and prominent variations in orientation. These findings provide further observational support for the emerging picture proposed by \citet{Fender_2025NatAs...9.1854F,fender2026linkobscuredaccretionmildly}, in which the fastest black hole jets propagate along fixed axes in unobscured systems, while heavily obscured systems preferentially launch slower and precessing jets. A summary of the observational differences between the 2025 jets and the pre-2019 relativistic jets is presented in \autoref{table:comparison}.

\subsection{Possible Origins of the Distinct Jet Properties} 
The physical origin of the obscured state in GRS\,1915+105 remains uncertain. Several possibilities have been proposed, including massive outflows or winds, changes in the outer accretion-disk geometry, and an increase in the accretion rate \citep[e.g.,][]{Miller_2020ApJ...904...30M,Neilsen_2020ApJ...902..152N,Koljonen_2020A&A...639A..13K,Sanchez-Sierras_2023A&A...680L..16S}.

As discussed by \citetalias{Yan_2026}, the large variations in jet orientation observed during the obscured state of GRS\,1915+105 could be explained by the presence of a warped, precessing accretion disk, as proposed by \citet{Miller_2025ApJ...995L..14M} based on recent X-ray spectroscopic observations with XRISM.

The relatively low jet speed may be related to an increase in the accretion rate of GRS\,1915+105. Early studies of correlated quasi-sinusoidal oscillations observed in the radio, infrared, and X-ray bands revealed characteristic periods of 20--60\,minutes \citep[e.g.,][]{Pooley_1997MNRAS.292..925P,Rodr_1997ApJ...474L.123R,Eikenberry_1998ApJ...494L..61E,Fender_1997MNRAS.290L..65F,Fender_1998MNRAS.300..573F,Fender_2000MNRAS.318L...1F,Mirabel_1998A&A...330L...9M}. In contrast, \citet{Rodr_2025ApJ...986..108R} recently reported a shorter period of only $\sim 8$\,minutes in observations obtained in 2023. Based on this change, they suggested that the semimajor axis of the accretion disk in 2023 may have decreased, probably implying an increase in the accretion rate by a factor of $\sim 4.3$.

In addition, \citet{fender2026linkobscuredaccretionmildly} discussed that the slower jets observed in obscured systems may result from rapid mass loading in dense environments, causing the jets to decelerate quickly shortly after launch. If the obscured state of GRS\,1915+105 is indeed associated with a denser environment and a higher accretion rate, such a mechanism could naturally account for the relatively low jet speed inferred in this work. Notably, {\tt JWST} observations in 2023 suggest a moderately high but still sub-Eddington accretion rate ($\sim$5--30\% of the Eddington limit) for GRS\,1915+105 \citep{Gandhi_2025MNRAS.537.1385G}. More recently, infrared and X-ray observations suggest an intrinsic weakening in the accretion activity of GRS\,1915+105 by 2--4 orders of magnitude (or more) around November 2025 or earlier \citep{gandhi2026intrinsicdeclineaccretionactivity}. Further multi-wavelength observations will be crucial for testing the scenarios proposed above.

\section{Summary} \label{sec:Summary}
Recently, \citet{Fender_2025NatAs...9.1854F,fender2026linkobscuredaccretionmildly} proposed a new paradigm in which jets launched by XRBs in unobscured and obscured states likely exhibit markedly different propagation properties. Since 2019, GRS\,1915+105 has transitioned from an unobscured state to an X-ray-obscured state, which continues to the present day. While the pre-2019 relativistic jets have been extensively studied, the properties of the post-2019 jets remain comparatively poorly constrained.

During strong radio flares in 2025, we captured two contrasting jet morphologies with the EAVN at 6.7\,GHz. The first epoch, obtained following a flare evolving from an optically thick to an optically thin spectrum, reveals a bright radio core accompanied by an extended jet structure. In contrast, the second epoch, observed near the peak of another flare exhibiting optically thin emission at lower frequencies, shows no detectable radio core and is dominated by two bright, symmetric, well-separated jet blobs. The hour-timescale light curves in both epochs reveal a gradual decline in flux density.

If the 2025 jets moved across the sky as fast as the pre-2019 superluminal jets, measurable radial displacements would be expected over the duration of our observations. However, no significant jet motion is detected in the time-binned VLBI images. This suggests that the 2025 jets do not exhibit the large apparent superluminal motions characteristic of the pre-2019 jets. Further support is provided by the inferred jet speed of $\beta\Gamma \lesssim 0.40$, which is lower than the values of $\beta\Gamma \gtrsim 1$ measured for the pre-2019 relativistic jets. 

Combined with the large variations in jet orientation observed since 2023 \citepalias{Yan_2026}, we suggest that jets launched during the obscured state of GRS\,1915+105 differ from the relativistic jets observed during the unobscured state, showing lower speeds and prominent variations in orientation (see \autoref{table:comparison}). These findings are consistent with the emerging paradigm proposed by \citet{Fender_2025NatAs...9.1854F,fender2026linkobscuredaccretionmildly}. We also briefly discuss possible explanations for the distinct jet properties observed during the current obscured state, including a warped, precessing accretion disk and an enhanced accretion rate.

%%%%%% %%%%%%
%% proof reading的时候，那个VLBI重点研发项目的基金号要改一下。可能要把 2024YFA1611500 改成2024YFA1611503 或 2024YFA1611504，到时候我再决定。
\begin{acknowledgments}
We sincerely thank the referee for the constructive comments, which improved the quality and clarity of the manuscript. We sincerely thank Dr.~Nobuyuki Sakai for providing the recalculation tables, which significantly facilitated this work. 
This work was supported by the National Key R\&D Program of China (grant Nos. 2024YFA1611500 and 2022SKA0120102).
X.Y. is supported by the China Postdoctoral Science Foundation under grant Nos. 2025M773200 and 2026T190850, and by the Xinjiang Tianchi Talent Program. 
L.C. acknowledges support from the Tianshan Talent Training Program (grant No. 2023TSYCCX0099). 
S.X. was supported by the National Research Council of Science \& Technology (NST) grant by the Korea government (MSIT; No. CAP22061-000).
W.J. is supported by the National Natural Science Foundation of China (grant Nos. 12173074 and 12573100). 
This work was partly supported by the Urumqi Nanshan Astronomy and Deep Space Exploration Observation and Research Station of Xinjiang (XJYWZ2303) and the Central Guidance for Local Science and Technology Development Fund (grant No. ZYYD2026JD01). 

The authors sincerely thank the EAVN coordinator, Dr.~Kiyoaki Wajima, for the prompt handling of our Target-of-Opportunity requests, as well as all EAVN staff members who rapidly initiated the follow-up observations and correlated the data after the 2025 radio flares were reported. 

Observations with the RATAN-600 telescope are supported by the Ministry of Science and Higher Education of the Russian Federation. The renovation of telescope equipment is currently provided within the national project ``Science and universities".

This work has made use of the East Asia VLBI Network (EAVN), which is operated under cooperative agreement by National Astronomical Observatory of Japan (NAOJ), Korea Astronomy and Space Science Institute (KASI), Shanghai Astronomical Observatory (SHAO), Xinjiang Astronomical Observatory (XAO), Yunnan Astronomical Observatory (YNAO), National Astronomical Research Institute of Thailand (Public Organization) (NARIT), and National Geographic Information Institute (NGII), with the operational support by Ibaraki University (for the operation of the Hitachi 32\,m and Takahagi 32\,m telescopes), Yamaguchi University (for the operation of the Yamaguchi 32\,m telescope), and Kagoshima University (for the operation of the VERA Iriki antenna).

\end{acknowledgments}

%% For this sample we use BibTeX plus aasjournalv7.bst to generate the

\bibliography{references}{}

@Inbook{Greisen2003,
author="Greisen, E. W.",
editor="Heck, Andr{\'e}",
title="AIPS, the VLA, and the VLBA",
bookTitle="Information Handling in Astronomy - Historical Vistas",
year="2003",
publisher="Springer Netherlands",
address="Dordrecht",
pages="109--125",
isbn="978-0-306-48080-5",
doi="10.1007/0-306-48080-8_7",
url="https://doi.org/10.1007/0-306-48080-8_7"
}

@INPROCEEDINGS{Shepherd_Difmap_1997ASPC..125...77S,
       author = {{Shepherd}, M.~C.},
        title = "{Difmap: an Interactive Program for Synthesis Imaging}",
    booktitle = {Astronomical Data Analysis Software and Systems VI},
         year = 1997,
       editor = {{Hunt}, Gareth and {Payne}, Harry},
       series = {Astronomical Society of the Pacific Conference Series},
       volume = {125},
        month = jan,
        pages = {77},
       adsurl = {https://ui.adsabs.harvard.edu/abs/1997ASPC..125...77S}
}

@ARTICLE{Fender_1999MNRAS.304..865F,
       author = {{Fender}, R.~P. and {Garrington}, S.~T. and {McKay}, D.~J. and {Muxlow}, T.~W.~B. and {Pooley}, G.~G. and {Spencer}, R.~E. and {Stirling}, A.~M. and {Waltman}, E.~B.},
        title = "{MERLIN observations of relativistic ejections from GRS 1915+105}",
      journal = {\mnras},
         year = 1999,
        month = apr,
       volume = {304},
       number = {4},
        pages = {865-876},
          doi = {10.1046/j.1365-8711.1999.02364.x},
archivePrefix = {arXiv},
       eprint = {astro-ph/9812150},
 primaryClass = {astro-ph},
       adsurl = {https://ui.adsabs.harvard.edu/abs/1999MNRAS.304..865F}
}

@ARTICLE{Rodr_1999ApJ...511..398R,
       author = {{Rodr{\'\i}guez}, L.~F. and {Mirabel}, I.~F.},
        title = "{Repeated Relativistic Ejections in GRS 1915+105}",
      journal = {\apj},
         year = 1999,
        month = jan,
       volume = {511},
       number = {1},
        pages = {398-404},
          doi = {10.1086/306642},
archivePrefix = {arXiv},
       eprint = {astro-ph/9808341},
 primaryClass = {astro-ph},
       adsurl = {https://ui.adsabs.harvard.edu/abs/1999ApJ...511..398R}
}

@ARTICLE{Mirabel_1994Natur.371...46M,
       author = {{Mirabel}, I.~F. and {Rodr{\'\i}guez}, L.~F.},
        title = "{A superluminal source in the Galaxy}",
      journal = {\nat},
         year = 1994,
        month = sep,
       volume = {371},
       number = {6492},
        pages = {46-48},
          doi = {10.1038/371046a0},
       adsurl = {https://ui.adsabs.harvard.edu/abs/1994Natur.371...46M}
}

@ARTICLE{Dhawan_2000ApJ...543..373D,
       author = {{Dhawan}, V. and {Mirabel}, I.~F. and {Rodr{\'\i}guez}, L.~F.},
        title = "{AU-Scale Synchrotron Jets and Superluminal Ejecta in GRS 1915+105}",
      journal = {\apj},
         year = 2000,
        month = nov,
       volume = {543},
       number = {1},
        pages = {373-385},
          doi = {10.1086/317088},
archivePrefix = {arXiv},
       eprint = {astro-ph/0006086},
 primaryClass = {astro-ph},
       adsurl = {https://ui.adsabs.harvard.edu/abs/2000ApJ...543..373D}
}

@ARTICLE{Miller-Jones_2005MNRAS.363..867M,
       author = {{Miller-Jones}, J.~C.~A. and {McCormick}, D.~G. and {Fender}, R.~P. and {Spencer}, R.~E. and {Muxlow}, T.~W.~B. and {Pooley}, G.~G.},
        title = "{Multiple relativistic outbursts of GRS1915+105: radio emission and internal shocks}",
      journal = {\mnras},
         year = 2005,
        month = nov,
       volume = {363},
       number = {3},
        pages = {867-881},
          doi = {10.1111/j.1365-2966.2005.09488.x},
archivePrefix = {arXiv},
       eprint = {astro-ph/0508230},
 primaryClass = {astro-ph},
       adsurl = {https://ui.adsabs.harvard.edu/abs/2005MNRAS.363..867M}
}

@ARTICLE{Miller-Jones_2007MNRAS.375.1087M,
       author = {{Miller-Jones}, J.~C.~A. and {Rupen}, M.~P. and {Fender}, R.~P. and {Rushton}, A. and {Pooley}, G.~G. and {Spencer}, R.~E.},
        title = "{Evidence for deceleration in the radio jets of GRS1915+105?}",
      journal = {\mnras},
         year = 2007,
        month = mar,
       volume = {375},
       number = {3},
        pages = {1087-1098},
          doi = {10.1111/j.1365-2966.2007.11381.x},
archivePrefix = {arXiv},
       eprint = {astro-ph/0612211},
 primaryClass = {astro-ph},
       adsurl = {https://ui.adsabs.harvard.edu/abs/2007MNRAS.375.1087M}
}

@ARTICLE{Fender_2004ARA&A..42..317F,
       author = {{Fender}, Rob and {Belloni}, Tomaso},
        title = "{GRS 1915+105 and the Disc-Jet Coupling in Accreting Black Hole Systems}",
      journal = {\araa},
         year = 2004,
        month = sep,
       volume = {42},
       number = {1},
        pages = {317-364},
          doi = {10.1146/annurev.astro.42.053102.134031},
archivePrefix = {arXiv},
       eprint = {astro-ph/0406483},
 primaryClass = {astro-ph},
       adsurl = {https://ui.adsabs.harvard.edu/abs/2004ARA&A..42..317F}
}

@article{Reid_2014,
doi = {10.1088/0004-637X/796/1/2},
url = {https://dx.doi.org/10.1088/0004-637X/796/1/2},
year = {2014},
month = {oct},
publisher = {The American Astronomical Society},
volume = {796},
number = {1},
pages = {2},
author = {M. J. Reid and J. E. McClintock and J. F. Steiner and D. Steeghs and R. A. Remillard and V. Dhawan and R. Narayan},
title = {A PARALLAX DISTANCE TO THE MICROQUASAR GRS 1915+105 AND A REVISED ESTIMATE OF ITS BLACK HOLE MASS},
journal = {The Astrophysical Journal}
}

@ARTICLE{Trushkin_2023ATel15964....1T,
       author = {{Trushkin}, S.~A. and {Nizhelskij}, N.~A. and {Tsybulev}, P.~G. and {Shevchenko}, A.~V.},
        title = "{New short-time radio and X-ray flare from GRS1915+105}",
      journal = {The Astronomer's Telegram},
         year = 2023,
        month = mar,
       volume = {15964},
        pages = {1},
        url  = {https://www.astronomerstelegram.org/?read=15964},
       adsurl = {https://ui.adsabs.harvard.edu/abs/2023ATel15964....1T}
}

@ARTICLE{Trushkin_2023ATel15974....1T,
       author = {{Trushkin}, S.~A. and {Nizhelskij}, N.~A. and {Tsybulev}, P.~G. and {Shevchenko}, A.~V.},
        title = "{Giant radio flare from GRS1915+105}",
      journal = {The Astronomer's Telegram},
         year = 2023,
        month = apr,
       volume = {15974},
        pages = {1},
        url = {https://www.astronomerstelegram.org/?read=15974},
       adsurl = {https://ui.adsabs.harvard.edu/abs/2023ATel15974....1T}
}

@ARTICLE{Trushkin_2023ATel16168....1T,
       author = {{Trushkin}, S.~A. and {Bursov}, N.~N. and {Nizhelskij}, N.~A. and {Tsybulev}, P.~G. and {Shevchenko}, A.~V.},
        title = "{New radio flare from GRS 1915+105}",
      journal = {The Astronomer's Telegram},
         year = 2023,
        month = aug,
       volume = {16168},
        pages = {1},
        url  = {https://www.astronomerstelegram.org/?read=16168},
       adsurl = {https://ui.adsabs.harvard.edu/abs/2023ATel16168....1T}
}

@ARTICLE{Motta_2021MNRAS.503..152M,
       author = {{Motta}, S.~E. and {Kajava}, J.~J.~E. and {Giustini}, M. and {Williams}, D.~R.~A. and {Del Santo}, M. and {Fender}, R. and {Green}, D.~A. and {Heywood}, I. and {Rhodes}, L. and {Segreto}, A. and {Sivakoff}, G. and {Woudt}, P.~A.},
        title = "{Observations of a radio-bright, X-ray obscured GRS 1915+105}",
      journal = {\mnras},
         year = 2021,
        month = may,
       volume = {503},
       number = {1},
        pages = {152-161},
          doi = {10.1093/mnras/stab511},
archivePrefix = {arXiv},
       eprint = {2101.01187},
 primaryClass = {astro-ph.HE},
       adsurl = {https://ui.adsabs.harvard.edu/abs/2021MNRAS.503..152M}
}

@ARTICLE{Reid_2023ApJ...959...85R,
       author = {{Reid}, M.~J. and {Miller-Jones}, J.~C.~A.},
        title = "{On the Distances to the X-Ray Binaries Cygnus X-3 and GRS 1915+105}",
      journal = {\apj},
         year = 2023,
        month = dec,
       volume = {959},
       number = {2},
          eid = {85},
        pages = {85},
          doi = {10.3847/1538-4357/acfe0c},
archivePrefix = {arXiv},
       eprint = {2309.15027},
 primaryClass = {astro-ph.HE},
       adsurl = {https://ui.adsabs.harvard.edu/abs/2023ApJ...959...85R}
}

@ARTICLE{Balakrishnan_2021ApJ...909...41B,
       author = {{Balakrishnan}, M. and {Miller}, J.~M. and {Reynolds}, M.~T. and {Kammoun}, E. and {Zoghbi}, A. and {Tetarenko}, B.~E.},
        title = "{The Novel Obscured State of the Stellar-mass Black Hole GRS 1915+105}",
      journal = {\apj},
         year = 2021,
        month = mar,
       volume = {909},
       number = {1},
          eid = {41},
        pages = {41},
          doi = {10.3847/1538-4357/abd6cb},
archivePrefix = {arXiv},
       eprint = {2012.15033},
 primaryClass = {astro-ph.HE},
       adsurl = {https://ui.adsabs.harvard.edu/abs/2021ApJ...909...41B}
}

@ARTICLE{Athulya_2023MNRAS.525..489A,
       author = {{Athulya}, M.~P. and {Nandi}, Anuj},
        title = "{Multimission view of the low-luminosity 'obscured' phase of GRS 1915+105}",
      journal = {\mnras},
         year = 2023,
        month = oct,
       volume = {525},
       number = {1},
        pages = {489-507},
          doi = {10.1093/mnras/stad2072},
archivePrefix = {arXiv},
       eprint = {2307.04206},
 primaryClass = {astro-ph.HE},
       adsurl = {https://ui.adsabs.harvard.edu/abs/2023MNRAS.525..489A}
}

@ARTICLE{Miller_2020ApJ...904...30M,
       author = {{Miller}, J.~M. and {Zoghbi}, A. and {Raymond}, J. and {Balakrishnan}, M. and {Brenneman}, L. and {Cackett}, E. and {Draghis}, P. and {Fabian}, A.~C. and {Gallo}, E. and {Kaastra}, J. and {Kallman}, T. and {Kammoun}, E. and {Motta}, S.~E. and {Proga}, D. and {Reynolds}, M.~T. and {Trueba}, N.},
        title = "{An Obscured, Seyfert 2-like State of the Stellar-mass Black Hole GRS 1915+105 Caused by Failed Disk Winds}",
      journal = {\apj},
         year = 2020,
        month = nov,
       volume = {904},
       number = {1},
          eid = {30},
        pages = {30},
          doi = {10.3847/1538-4357/abbb31},
archivePrefix = {arXiv},
       eprint = {2007.07005},
 primaryClass = {astro-ph.HE},
       adsurl = {https://ui.adsabs.harvard.edu/abs/2020ApJ...904...30M}
}

@ARTICLE{Gandhi_2025MNRAS.537.1385G,
       author = {{Gandhi}, P. and {Borowski}, E.~S. and {Byrom}, J. and {Hynes}, R.~I. and {Maccarone}, T.~J. and {Shaw}, A.~W. and {Adegoke}, O.~K. and {Altamirano}, D. and {Baglio}, M.~C. and {Bhargava}, Y. and {Britt}, C.~T. and {Buckley}, D.~A.~H. and {Buisson}, D.~J.~K. and {Casella}, P. and {Segura}, N. Castro and {Charles}, P.~A. and {Corral-Santana}, J.~M. and {Dhillon}, V.~S. and {Fender}, R. and {G{\'u}rpide}, A. and {Heinke}, C.~O. and {Igl}, A.~B. and {Knigge}, C. and {Markoff}, S. and {Mastroserio}, G. and {McCollough}, M.~L. and {Middleton}, M. and {Miller}, J.~M. and {Miller-Jones}, J.~C.~A. and {Motta}, S.~E. and {Paice}, J.~A. and {Pawar}, D.~D. and {Plotkin}, R.~M. and {Pradhan}, P. and {Ressler}, M.~E. and {Russell}, D.~M. and {Russell}, T.~D. and {Santos-Sanz}, P. and {Shahbaz}, T. and {Sivakoff}, G.~R. and {Steeghs}, D. and {Tetarenko}, A.~J. and {Tomsick}, J.~A. and {Vincentelli}, F.~M. and {George}, M. and {Gurwell}, M. and {Rao}, R. and {JWST Timing Consortium}},
        title = "{Rapid mid-infrared spectral timing with JWST: GRS 1915+105 during an MIR-bright and X-ray-obscured state}",
      journal = {\mnras},
         year = 2025,
        month = feb,
       volume = {537},
       number = {2},
        pages = {1385-1403},
          doi = {10.1093/mnras/staf036},
       adsurl = {https://ui.adsabs.harvard.edu/abs/2025MNRAS.537.1385G}
}

@ARTICLE{Pooley_1997MNRAS.292..925P,
       author = {{Pooley}, G.~G. and {Fender}, R.~P.},
        title = "{The variable radio emission from GRS 1915+=105}",
      journal = {\mnras},
         year = 1997,
        month = dec,
       volume = {292},
       number = {4},
        pages = {925-933},
          doi = {10.1093/mnras/292.4.925},
archivePrefix = {arXiv},
       eprint = {astro-ph/9708171},
 primaryClass = {astro-ph},
       adsurl = {https://ui.adsabs.harvard.edu/abs/1997MNRAS.292..925P}
}

@ARTICLE{Sanchez-Sierras_2023A&A...680L..16S,
       author = {{S{\'a}nchez-Sierras}, J. and {Mu{\~n}oz-Darias}, T. and {Motta}, S.~E. and {Fender}, R.~P. and {Bahramian}, A. and {Mart{\'\i}nez-Sebasti{\'a}n}, C. and {Fern{\'a}ndez-Ontiveros}, J.~A. and {Casares}, J. and {Armas Padilla}, M. and {Green}, D.~A. and {Mata S{\'a}nchez}, D. and {Strader}, J. and {Torres}, M.~A.~P.},
        title = "{Fast infrared winds during the radio-loud and X-ray obscured stages of the black hole transient GRS 1915+105}",
      journal = {\aap},
         year = 2023,
        month = dec,
       volume = {680},
          eid = {L16},
        pages = {L16},
          doi = {10.1051/0004-6361/202348184},
archivePrefix = {arXiv},
       eprint = {2311.12933},
 primaryClass = {astro-ph.HE},
       adsurl = {https://ui.adsabs.harvard.edu/abs/2023A&A...680L..16S}
}

@ARTICLE{Trushkin_2025ATel16976....1T,
       author = {{Trushkin}, S.~A. and {Tsybulev}, P.~G. and {Shevchenko}, A.~V. and {Bursov}, N.~N. and {Nizhelskij}, N.~A.},
        title = "{The bright radio flare from microquasar GRS 1915+105}",
      journal = {The Astronomer's Telegram},
         year = 2025,
        month = jan,
       volume = {16976},
        pages = {1},
       adsurl = {https://ui.adsabs.harvard.edu/abs/2025ATel16976....1T}
}

@ARTICLE{Mirabel_1999ARA&A..37..409M,
       author = {{Mirabel}, I.~F. and {Rodr{\'\i}guez}, L.~F.},
        title = "{Sources of Relativistic Jets in the Galaxy}",
      journal = {\araa},
         year = 1999,
        month = jan,
       volume = {37},
        pages = {409-443},
          doi = {10.1146/annurev.astro.37.1.409},
archivePrefix = {arXiv},
       eprint = {astro-ph/9902062},
 primaryClass = {astro-ph},
       adsurl = {https://ui.adsabs.harvard.edu/abs/1999ARA&A..37..409M}
}

@ARTICLE{Rodr_1997ApJ...474L.123R,
       author = {{Rodr{\'\i}guez}, Luis F. and {Mirabel}, I. F{\'e}lix},
        title = "{Fast Sinusoidal Oscillations in the Radio Flux of GRS 1915+105}",
      journal = {\apjl},
         year = 1997,
        month = jan,
       volume = {474},
       number = {2},
        pages = {L123-L125},
          doi = {10.1086/310443},
       adsurl = {https://ui.adsabs.harvard.edu/abs/1997ApJ...474L.123R}
}

@ARTICLE{Mirabel_1998A&A...330L...9M,
       author = {{Mirabel}, I.~F. and {Dhawan}, V. and {Chaty}, S. and {Rodriguez}, L.~F. and {Marti}, J. and {Robinson}, C.~R. and {Swank}, J. and {Geballe}, T.},
        title = "{Accretion instabilities and jet formation in GRS 1915+105}",
      journal = {\aap},
         year = 1998,
        month = feb,
       volume = {330},
        pages = {L9-L12},
          doi = {10.48550/arXiv.astro-ph/9711097},
archivePrefix = {arXiv},
       eprint = {astro-ph/9711097},
 primaryClass = {astro-ph},
       adsurl = {https://ui.adsabs.harvard.edu/abs/1998A&A...330L...9M}
}

@ARTICLE{Fender_1997MNRAS.290L..65F,
       author = {{Fender}, R.~P. and {Pooley}, G.~G. and {Brocksopp}, C. and {Newell}, S.~J.},
        title = "{Rapid infrared flares in GRS 1915+105: evidence for infrared synchrotron emission}",
      journal = {\mnras},
         year = 1997,
        month = oct,
       volume = {290},
       number = {4},
        pages = {L65-L69},
          doi = {10.1093/mnras/290.4.L65},
archivePrefix = {arXiv},
       eprint = {astro-ph/9707317},
 primaryClass = {astro-ph},
       adsurl = {https://ui.adsabs.harvard.edu/abs/1997MNRAS.290L..65F}
}

@ARTICLE{Fender_1998MNRAS.300..573F,
       author = {{Fender}, R.~P. and {Pooley}, G.~G.},
        title = "{Infrared synchrotron oscillations in GRS 1915+105}",
      journal = {\mnras},
         year = 1998,
        month = oct,
       volume = {300},
       number = {2},
        pages = {573-576},
          doi = {10.1046/j.1365-8711.1998.01921.x},
archivePrefix = {arXiv},
       eprint = {astro-ph/9806073},
 primaryClass = {astro-ph},
       adsurl = {https://ui.adsabs.harvard.edu/abs/1998MNRAS.300..573F}
}

@ARTICLE{Fender_2000MNRAS.318L...1F,
       author = {{Fender}, R.~P. and {Pooley}, G.~G.},
        title = "{Giant repeated ejections from GRS 1915+105}",
      journal = {\mnras},
         year = 2000,
        month = oct,
       volume = {318},
       number = {1},
        pages = {L1-L5},
          doi = {10.1046/j.1365-8711.2000.03847.x},
archivePrefix = {arXiv},
       eprint = {astro-ph/0006278},
 primaryClass = {astro-ph},
       adsurl = {https://ui.adsabs.harvard.edu/abs/2000MNRAS.318L...1F}
}

@ARTICLE{Eikenberry_1998ApJ...494L..61E,
       author = {{Eikenberry}, Stephen S. and {Matthews}, Keith and {Morgan}, Edward H. and {Remillard}, Ronald A. and {Nelson}, Robert W.},
        title = "{Evidence for a Disk-Jet Interaction in the Microquasar GRS 1915+105}",
      journal = {\apjl},
         year = 1998,
        month = feb,
       volume = {494},
       number = {1},
        pages = {L61-L64},
          doi = {10.1086/311158},
archivePrefix = {arXiv},
       eprint = {astro-ph/9710374},
 primaryClass = {astro-ph},
       adsurl = {https://ui.adsabs.harvard.edu/abs/1998ApJ...494L..61E}
}

@ARTICLE{Rodr_2025ApJ...986..108R,
       author = {{Rodr{\'\i}guez}, Luis F. and {Mirabel}, I. F{\'e}lix},
        title = "{An Unusual Change in the Radio Jets of GRS 1915+105}",
      journal = {\apj},
         year = 2025,
        month = jun,
       volume = {986},
       number = {1},
          eid = {108},
        pages = {108},
          doi = {10.3847/1538-4357/adda33},
archivePrefix = {arXiv},
       eprint = {2503.01105},
 primaryClass = {astro-ph.HE},
       adsurl = {https://ui.adsabs.harvard.edu/abs/2025ApJ...986..108R}
}

@ARTICLE{Miller_2025ApJ...995L..14M,
       author = {{Miller}, Jon M. and {Gu}, Liyi and {Raymond}, John and {Brenneman}, Laura and {Gallo}, Elena and {Gandhi}, Poshak and {Kallman}, Timothy and {Kobayashi}, Shogo and {Mao}, Junjie and {Rogantini}, Daniele and {Shidatsu}, Megumi and {Ueda}, Yoshihiro and {Xiang}, Xin and {Zoghbi}, Abderahmen},
        title = "{XRISM Spectroscopy of the Stellar-mass Black Hole GRS 1915+105}",
      journal = {\apjl},
         year = 2025,
        month = dec,
       volume = {995},
       number = {1},
          eid = {L14},
        pages = {L14},
          doi = {10.3847/2041-8213/ae2123},
archivePrefix = {arXiv},
       eprint = {2510.25089},
 primaryClass = {astro-ph.HE},
       adsurl = {https://ui.adsabs.harvard.edu/abs/2025ApJ...995L..14M}
}

@ARTICLE{Trushkin_2000A&AT...19..525T,
       author = {{Trushkin}, S.~A.},
        title = "{Radio emission of galactic X-ray binaries with relativistic jets}",
      journal = {Astronomical and Astrophysical Transactions},
         year = 2000,
        month = jan,
       volume = {19},
       number = {3},
        pages = {525-535},
          doi = {10.1080/10556790008238598},
archivePrefix = {arXiv},
       eprint = {astro-ph/0001033},
 primaryClass = {astro-ph},
       adsurl = {https://ui.adsabs.harvard.edu/abs/2000A&AT...19..525T}
}

@INPROCEEDINGS{Trushkin_2008mqw..confE..32T,
       author = {{Trushkin}, Sergei A. and {Nizhelskij}, Nikolaj A. and {Bursov}, Nikolaj N.},
        title = "{Longterm multi-frequency monitoring of microquasars}",
    booktitle = {Microquasars and Beyond},
         year = 2008,
        month = jan,
          eid = {32},
        pages = {32},
          doi = {10.22323/1.062.0032},
archivePrefix = {arXiv},
       eprint = {0810.3376},
 primaryClass = {astro-ph},
       adsurl = {https://ui.adsabs.harvard.edu/abs/2008mqw..confE..32T}
}

@ARTICLE{Fender_2025NatAs...9.1854F,
       author = {{Fender}, R.~P. and {Motta}, S.~E.},
        title = "{The connection between the fastest astrophysical jets and the spin axis of their black hole}",
      journal = {Nature Astronomy},
         year = 2025,
        month = dec,
       volume = {9},
        pages = {1854-1859},
          doi = {10.1038/s41550-025-02665-w},
       adsurl = {https://ui.adsabs.harvard.edu/abs/2025NatAs...9.1854F}
}

@article{fender2026linkobscuredaccretionmildly,
    author = {{Fender}, R.~P. and {Motta}, S.~E.},
    title = {The link between obscured accretion and mildly relativistic precessing jets},
    journal = {MNRAS},
    pages = {stag1112},
    year = {2026},
    month = {06},
    issn = {0035-8711},
    doi = {10.1093/mnras/stag1112},
    url = {https://doi.org/10.1093/mnras/stag1112},
    eprint = {https://academic.oup.com/mnras/advance-article-pdf/doi/10.1093/mnras/stag1112/68590135/stag1112.pdf},
}

@ARTICLE{Yan_2026,
       author = {{Yan}, Xi and {Cui}, Lang and {Jiang}, Wu and {Yan}, Zhen and {Frey}, S{\'a}ndor and {Trushkin}, Sergei and {Mufakharov}, Timur and {Dhaka}, Ruchika and {Xu}, Shuangjing},
        title = "{Discovery of Unusual Jet Orientation Variations in the Microquasar GRS 1915+105}",
      journal = {\apjl},
         year = 2026,
        month = jul,
       volume = {1006},
       number = {1},
          eid = {L15},
        pages = {L15},
          doi = {10.3847/2041-8213/ae80c8},
archivePrefix = {arXiv},
       eprint = {2606.16228},
 primaryClass = {astro-ph.HE},
       adsurl = {https://ui.adsabs.harvard.edu/abs/2026ApJ..1006L..15Y}
}

@ARTICLE{Jiang_2026ApJ..1000L..45J,
       author = {{Jiang}, Wu and {Yan}, Xi and {Yan}, Zhen and {Li}, Ya-Ping and {Cui}, Lang and {Shen}, Zhi-Qiang},
        title = "{A Large Misalignment between Continuous Jet and Discrete Ejecta in Microquasar GRS 1915+105 during Its Obscured Phase}",
      journal = {\apjl},
         year = 2026,
        month = apr,
       volume = {1000},
       number = {2},
          eid = {L45},
        pages = {L45},
          doi = {10.3847/2041-8213/ae5186},
archivePrefix = {arXiv},
       eprint = {2604.00357},
 primaryClass = {astro-ph.HE},
       adsurl = {https://ui.adsabs.harvard.edu/abs/2026ApJ..1000L..45J}
}

@ARTICLE{Neilsen_2020ApJ...902..152N,
       author = {{Neilsen}, J. and {Homan}, J. and {Steiner}, J.~F. and {Marcel}, G. and {Cackett}, E. and {Remillard}, R.~A. and {Gendreau}, K.},
        title = "{A NICER View of a Highly Absorbed Flare in GRS 1915+105}",
      journal = {\apj},
         year = 2020,
        month = oct,
       volume = {902},
       number = {2},
          eid = {152},
        pages = {152},
          doi = {10.3847/1538-4357/abb598},
archivePrefix = {arXiv},
       eprint = {2010.14512},
 primaryClass = {astro-ph.HE},
       adsurl = {https://ui.adsabs.harvard.edu/abs/2020ApJ...902..152N}
}

@ARTICLE{Koljonen_2020A&A...639A..13K,
       author = {{Koljonen}, K.~I.~I. and {Tomsick}, J.~A.},
        title = "{The obscured X-ray binaries V404 Cyg, Cyg X-3, V4641 Sgr, and GRS 1915+105}",
      journal = {\aap},
         year = 2020,
        month = jul,
       volume = {639},
          eid = {A13},
        pages = {A13},
          doi = {10.1051/0004-6361/202037882},
archivePrefix = {arXiv},
       eprint = {2004.08536},
 primaryClass = {astro-ph.HE},
       adsurl = {https://ui.adsabs.harvard.edu/abs/2020A&A...639A..13K}
}

@ARTICLE{Motta_2026ATel17865....1M,
       author = {{Motta}, S. and {Fender}, R. and {X-KAT Collaboration} and {Marino}, A. and {Carotenuto}, F. and {Baglio}, M.~C. and {van den Eijnden}, J. and {Atri}, P. and {Williams-Baldwin}, D.},
        title = "{MeerKAT observes the radio fading of GRS 1915+105 to its lowest luminosity level since discovery}",
      journal = {The Astronomer's Telegram},
         year = 2026,
        month = jul,
       volume = {17865},
        pages = {1},
       adsurl = {https://ui.adsabs.harvard.edu/abs/2026ATel17865....1M}
}

@misc{gandhi2026intrinsicdeclineaccretionactivity,
      title={An intrinsic decline of accretion activity in GRS 1915+105}, 
      author={Poshak Gandhi and Peter G. Boorman},
      year={2026},
      eprint={2607.06200},
      archivePrefix={arXiv},
      primaryClass={astro-ph.HE},
      url={https://arxiv.org/abs/2607.06200}, 
}
\bibliographystyle{aasjournalv7}
%\bibliographystyle{mnras}



\end{document}